\documentclass{elsart}
\usepackage{amssymb}
\usepackage{graphicx}
\usepackage{dcolumn}
\usepackage{bm}
\usepackage{here}
\usepackage{tabularx}

\begin{document}
\begin{frontmatter}
\title{Quantum Diffusion in Polaron Model of poly(dG)-poly(dC) and
poly(dA)-poly(dT) DNA polymers 
}
\author[Yamada]{Hiroaki Yamada \corauthref{cor}}
\author[Starikov]{, Eugen B. Starikov}
\author[Hennig]{, Dirk Hennig}
\corauth[cor]{Corresponding author. 
electronic mail address: hyamada@uranus.dti.ne.jp, 
Fax: +81-25-267-1941.
}

\address[Yamada]{Yamada Physics Research Laboratory, 5-7-14 Aoyama, Niigata 950-2002, Japan}
\address[Starikov]{
Institut f$\ddot{u}$r Nanotechnologie
Forschungszentrum Karlsruhe
Postfach3640, 70621 Karlsruhe,  Germany
}
\address[Hennig]{
Humboldt Universit$\ddot{a}$t zu Berlin, Institut f\"ur Physik, Newtonstr. 15, 12489 Berlin,
Germany}


\begin{abstract}
We numerically investigate quantum diffusion of an electron in a
model of poly(dG)-poly(dC) and poly(dA)-poly(dT) DNA
polymers with fluctuation of the parameters due to the impact of colored noise. 
The randomness is introduced by fluctuations of distance between two consecutive bases
along the stacked base pairs.
We demonstrate that in the model the decay time of the correlation can control the spread of the 
electronic wavepacket. 
Furthermore it is shown that in a motional narrowing regime the averaging over fluctuation 
causes ballistic propagation of the wavepacket,
and in the adiabatic regime the electronic states are affected by localization.
\end{abstract}

\end{frontmatter}

\section{Introduction}
\label{sec01}
Charge transfer/transport properties in DNA attract lively interest
among physicists, chemists and engineers 
\cite{porath04,tran00,campbell04,hermon98,lebard03,roche03,yamada04,malyshev07}.
It is now well established that diverse DNA structural
deformations are of extreme importance during the charge transfer/transport process,
because they help to create polarons which promote not only the formation of a localized electronic state,
but may also assist in rendering the latter mobile \cite{conwell00}.
Remarkably, recent sophisticated experimental techniques, like, for example, spin-echo spectroscopy,
allow to measure stochastic structural dynamics of monomers
in polymer chains, such as double-stranded DNA \cite{xray,shusterman04,bern76}:
There is a wealth of dynamical modes possessed of a quasi-continuum spectrum. 
In principle, each of these can influence DNA charge transfer/transport,
but, since there are more or less active modes \cite{starikov05}, it is possible to
take the whole manifold of DNA molecular vibrations into two parts - those which are most active, plus a 
"stochastic bath" consisting of all other ones.
For the present, a number of the polaron models have
been proposed to describe charge transfer/transport 
in DNA polymers, see, for example, \cite{conwell00,conwell03,hennig02,palmero04,chang04,kats02}. 
On the other hand, computer simulations have pinpointed the crucial significance of
dynamical disorder for DNA transfer/transport \cite{troisi02,tanaka03,sankey03,voityuk04,tanaka06}, 
and several attempts to formulate stochastical models for the interplay of the former and the latter 
have already appeared in the literature, see, for example, \cite{baranovskii05,sakamoto07}.

In this communication, we shall deal with the polaron-like model by Hennig 
and coworkers as described in the works \cite{hennig02,palmero04}, 
where charge+breather propagation along DNA homopolynucleotide duplexes, i.e.
in the both poly(dG)-poly(dC) and poly(dA)-poly(dT) DNA polymers
has been studied. These works find that the coupled motion of
charges and breathers connected with localized structural vibrations may contribute to highly efficient long-range
conductivity.

In our previous paper, we have investigated localization properties of
electronic states in the adiabatic limit using a
stochastic-bond-vibration approach for poly(dG)-poly(dC) and
poly(dA)-poly(dT) DNA polymers within the framework of the polaron model \cite{yamada05}. 
That time, we assumed that the disorder is
caused by DNA vibrational modes, and it influences the charge transfer/transport along DNA
duplexes via electron-vibrational coupling. 
Here we present numerical results concerning the
influence of stochastic changes in DNA hydrogen-bond stretchings and double-helical twisting
angles, as well as the effects of finite system size, on the
electron localization properties. 

Specifically, in the present paper 
we numerically investigate quantum diffusion of electron in the Hennig
 model of poly(dG)-poly(dC) and poly(dA)-poly(dT) with added fluctuations.   
The latter are described by a colored noise associated with stochastic dynamics 
of the distances $r(t)$ between two Watson-Crick base pair partners:
 $ <r(t)r(t^{'})> = r_{0}^2 \exp(-|t-t^{'}|/\tau)$.
These fluctuations can be regarded as a stochastic process at high temperature, 
with phonon modes being randomly excited.
In the model the characteristic decay time $\tau$ of the correlation can control the spread of the 
electronic wavepacket. 
Interestingly, the white-noise limit $\tau \to 0$ can in effect correspond to 
a sort of motional narrowing regime,
(see, for example, \cite{berthelot06}) 
because we find that such a regime causes ballistic propagation of 
the wavepacket through homogeneous DNA duplexes.
Still, in the adiabatic limit $\tau \to \infty$, DNA electronic states 
should be strongly affected by localization.

 The amplitude $r_0$ of the random fluctuations within the base
 pairs (the fluctuation of the distance between two bases in a base pair) and
 the correlation time $\tau$,
are very critical parameters for the diffusive properties of wavepackets. 
 The finding of ballistic behavior in the white-noise limit vs. localization in the
 adiabatic limit is interesting, since there is
 a number of experimental works \cite{tao04,porath05,zalinge06} 
observing ballistic conductance of DNA in water solutions, which is also
 temperature-independent \cite{zalinge06}. Zalinge {\it et al.} 
have tried to 
explain the latter effect, using a kind
 of acoustic phonon motions in DNA duplexes \cite{zalinge06}, 
which seems to be plausible \cite{mandal06},
 but not the only possible physical reason.  
We will propose an alternative explanation for the observed temperature-independent conductance, based
upon our numerical results. 

The outline of the present paper is as follows. In the next section we
introduce our DNA model for the investigation of
its diffusive electronic properties.
In Sect. 3, we give a brief explanation for the characteristic motion of wavepackets 
under the impact of colored noise.
In Sec. 4 we present numerical results concerning the
influence of the hydrogen-bond stretching fluctuations and twist
angles on the localization properties. Furthermore we comment on the  
relation between our numerical result and the experimental one.
The last section contains our conclusions.

\section{Model and parameters}
The Hamiltonian
for the electronic part in our DNA model is given by
 \begin{eqnarray}
H_{el} & = & \sum_{n}E_{n}C_{n}^{\dagger}C_{n} 
- \sum_{n}V_{nn+1}(C_{n}^{\dagger}C_{n+1}+C_{n+1}C_{n}^{\dagger}),
\end{eqnarray}
 where $C_{n}$ and $C_{n}^{\dagger}$ are creation and annihilation
operators of an electron at the site $n$. The on-site energies $E_{n}$
are represented as
\begin{eqnarray}
E_{n} & = & E_{0}+kr_{n},
\end{eqnarray}
 where $E_{0}$ is a constant and $r_{n}$ denotes the structural fluctuation
caused by the coupling with the transversal Watson-Crick H-bonding stretching
vibration.

The transfer integral $V_{nn+1}$ depends on the three-dimensional
 distance $d_{nn+1}$ between adjacent stacked base pairs, labeled by $n$ and
$n+1$, along each strand - and is expressed as follows,
\begin{eqnarray}
V_{nn+1} & = & V_{0}(1-\alpha d_{nn+1})\,.
\end{eqnarray}
The parameters $k$ and $\alpha$ describe the strength of
the interaction between the electronic and vibrational variables.
The 3D displacements $d_{nn+1}$  bring about also a variation of the distances
between neighboring bases along each strand.  The
 first order Taylor expansion around the equilibrium positions is given by
\begin{eqnarray}
d_{nn+1} & = & \frac{R_{0}}{\ell_{0}}(1-\cos\theta_{0})(r_{n}+r_{n+1}).
\end{eqnarray}
 $R_{0}$ represents the equilibrium radius of the helix, $\theta_{0}$
is the equilibrium double-helical twist angle between base pairs, and $\ell_{0}$
the equilibrium distance between bases along one strand given by
\begin{eqnarray}
\ell_{0} & = & (a^{2}+4R_{0}^{2}\sin^{2}(\theta_{0}/2))^{1/2},
\end{eqnarray}
 with $a$ being the distance between neighboring base pairs in the
direction of the helix axis. 
We adopt realistic values of the parameters obtained from
the semi-empirical quantum-chemical calculations \cite{hennig04}. (See table 1.)

\begin{table}[htb]
\begin{center}
\begin{tabularx}{70mm}{XX}  \hline \hline
 {\em parameter} & {\em value} \\ \hline
 $E_0$    & $0.1[eV]$   \\
$V_0$     & $0.1[eV]$  \\
 $a$  & $3.4 [$\AA$]$ \\
  $R_0$   & $10 [$\AA$]$  \\
 $\theta_0$     & $36[^{\circ}]$  \\  
$k_{AT}$ & $0.778917 [eV$\AA$^{-1}]$ \\
$\alpha_{AT}$ &    $0.053835 [$\AA$^{-1}]$  \\
$k_{GC}$  & $-0.090325 [eV$\AA$^{-1}]$ \\
$\alpha_{GC}$ & $0.383333 [$\AA$^{-1}]$ \\
 \hline\hline
\end{tabularx}
\end{center}
\caption{ Basic parameters for DNA molecules. 
The subscripts, $AT$ and $GC$, for $k$ and $\alpha$ denote 
for ones of the poly(dA)-poly(dT) and poly(dG)-poly(dC) DNA polymers, respectively. 
}
\end{table}

The 
Schr\"{o}dinger equation 
describing the temporal evolution of the electron state vector $|\psi >$ reads 
\begin{eqnarray}
i\hbar \frac{ \partial |\psi > }{\partial t} = H_{el}(t) |\psi >.
\end{eqnarray}
The explicit time-dependence of the Hamiltonian $H_{el}(t)$ is given through both the time-dependence
of the on-site and hopping terms. 
The equation can be expressed by scaled dimensionless variables 
 as,
\begin{eqnarray}
i \hbar_{eff} \frac{ \partial \phi_n }{\partial t} & = & E_n(t) \phi_n -V_{nn+1}(t) \phi_{n+1} -V_{n-1n}(t) \phi_{n-1} ,  
\end{eqnarray}
where $\phi_n =<n|\psi>$ and the effective Planck constant $\hbar_{eff}=0.53$.
We redefined the scaled dimensionless variables
 $E_n(t)$ and $V_{n n+1}(t)$ in Eq.(1), as 
 $\frac{E_n}{V_0} \to E_n$,  $\frac{V_{nn+1}}{V_0} \to V_{nn+1}$.  

In addition to the DNA homopolymer duplexes, we also investigate the
mixed sequence consisting of two types
of the Watson-Crick pairs. Then, as a zero-order approximation,
the electron-phonon coupling parameters for the mixed GC/AT stacks
are taken here to be equal to the values obtained for
poly(dG)-poly(dC) and poly(dA)-poly(dT) DNA polymers.

We used mainly 4th order Runge-Kutta-Gill method 
in the numerical simulation for the time evolution with
 time step $\delta t=0.01$.  
In some cases we confirmed the accuracy of the so obtained results  by complete accord with the results gained with the help of 
a 6th order symplectic integrator that is higher order unitary integrations.

\section{Fluctuation of the $r_n(t)$ and motion of wavepackets}
The explicit time-dependence of the Hamiltonian comes from the fluctuation of the variable $r_n(t)$
at each site $n$. To mimic these fluctuations we use a  
Gaussian-Markovian process with standard deviation $r_{n0}$(amplitude) 
and correlation time $\tau$ 
characterized by the following covariance
\begin{eqnarray}
 C(t-t^{'}) & \equiv & <r_n(t)r_m(t^{'})> = \delta_{nm} r_{n0}^2 e^{-|t-t^{'}| / \tau }, 
\end{eqnarray}
where  $<r_n(t)>= 0$. 
This process is called the Ornstein-Uhlenbeck process. There is no spatial correlation, viz, the stochastic fluctuations at different sites are 
independent of each other. The white-noise limit corresponds to $\tau \to 0$. 
On the other hand, the adiabatic limit ($\tau \to \infty $)
corresponds static Anderson model. 
A numerical way for the  generation of the colored noise is given in appendix A.
The fluctuations are characterized by two parameters, i.e. 
amplitude $r_{n0}$ and  correlation time $\tau$. 
Another relevant quantity is, 
 \begin{eqnarray}
D_0 &\equiv & \int_0^\infty C(t) dt  =  r_{n0}^2  \tau\,.
\end{eqnarray}
The quantity $D_0$ expresses  total strength for the random fluctuation 
and is related to self-diffusion coefficient \cite{sumi77,inaba81,ezaki91}.
The limit $\tau \to 0$ and $r_{n0} \to \infty$
with keeping $D_0=const.$ yields Gaussian white-noise characterized by $<r_n(t)r_n(t^{'})> = 2D_0 \delta(t-t^{'}) $. 
For $|t-t^{'}|  >  \tau $, we may treat $r_n(t)$ and $r_n(t^{'})$ as statistically independent 
quantities. This limiting case becomes valid when the lattice temperature is well above the 
Debye temperature. The characteristic decay rate $\tau_c$ of the correlation function is on the order 
of the frequency $\Omega$ of the slow bond vibrations due to the relation 
\begin{eqnarray}
\tau_c &=& \frac{\hbar \Omega}{V_0}   \sim   0.05.
\end{eqnarray}
The adiabatic limit holds true for $\tau >> \tau_c$ allowing for the application of the Born-Oppenheimer approximation. 

In general, in our numerical simulation, the order of mean values $<E_n(t)>, <V_{nn+1}(t)>$ and 
the amplitude of the fluctuations (standard deviation) $\Delta E_n(t),  \Delta V_{nn+1}(t)$  
 of the  on-site energy and the transfer energy
are  estimated as follows: $<E_n(t)> \sim O(1),  \Delta E_n(t)  \sim  k  r_{n0}, 
<V_{nn+1}(t)> \sim O(1),  \Delta V_{nn+1}(t)  \sim  2 \alpha r_{n0}F_0$, 
where $F_{0} \equiv \frac{R_0}{\ell_0} (1-\cos \theta_0) $. 
As the fluctuating amplitude is concerned it can be naturally incorporated by assuming the temperature dependence as,
\begin{eqnarray}
 r^2_{n0}  & \propto &  k_B T, 
\end{eqnarray}
Particularly, at room temperature, i.e. for  $ k_B T  =  0.026 [eV]$, 
 the value of the scaled dimensionless thermal energy is given by $ k_B T/V_0  =  0.26$.

The motion of the electron in the fluctuating medium is crucially influenced by the values of the parameters $r_{n0}$ and $\tau$. 
Another important parameter is the band width $B$ of the electron system 
which is given by the static case without fluctuation $r_{n0}=0$.  In the scaled Hennig model for 
DNA the band width is $B \sim 2$, although the exact value depends on A-T or G-C or mixed models.
In order to infer on the onsequences of the noise for the elelctronmotion it illustratitive to
express the parameters of the noise in units of $r_{n0}$. 
Thus, with regard to its influence on the motion the noise is quantified by the effective parameters  
$B/r_{n0}$ and $1/(\tau r_{n0})$. 
While  $B$ and $1/\tau$ have the effect of narrowing the absorption linewidth, 
$r_{n0}$ has the oppsite effect, namely broadening of the linewidth. 
The motional narrowing of the resonant absorption linewidth becomes important 
when the fluctuation rate is larger than the amplitude, i.e. $\tau^{-1} > r_{n0}$.
It then follows that time-dependent perturbation theory is true for  $B/r_{n0}>>1$ or $1/(\tau r_{n0}) >>1$, 
and the adiabatic approximation is applicable in the region $\tau B >>1$.
In the following sections we show some typical wavepacket dynamics in the different parameter regimes.

Note that the ballistic propagation of a wave packet occurs  when $\tau \leq  \delta t(=0.01)$,  
i.e. the fluctuation is very rapid and the system is reduced to a regular system without any 
disorder.

\section{Numerical results}
\label{sec03}
In this section, we show the numerical results of the wavepacket dynamics.
We consider quantum diffusion of an initially localized wavepacket $\phi_n(t=0)=  \delta_{n n_0}$.
Then we  mainly monitor the time-dependence of the mean square displacement (MSD) 
\begin{eqnarray}
 m(t) &= & \sum_n^N <\phi_n |(\hat{n}-n_0)^2| \phi_n>,  
\end{eqnarray}
and the distribution function $P(n,t)\equiv |<n|\psi (t) >| =|\phi_n(t)| $.
We used $N=2^{12}$ and $n_0=N/2$ through this paper.

\subsection{Constant hopping term $V_{n n+1}=1(\alpha=0)$ }

First we present our numerical results of simple cases with a constant transfer integral $V_{nn+1}=1.0$
 for all $n$. We illustrate the typical motion of a wavepacket in cases when only the on-site energy $E_n(t)$ fluctuates according to a  
Ornstein-Uhlenbeck process.
Although we used mixed model in this case the qualitative result does not depend on the 
type of model even if  we adapt A-T model and G-C model. 

Figure 1(a)  shows time-dependence of the MSD for various correlation times 
$\tau=10^5, 1, 0.01$ at $r_{n0}=1.0$. In  
Fig.1(b) and (c)  we depict some snap-shots of $|\phi_n(t)| $ in cases with $\tau=0.01$ and 
$\tau=10^5$, respectively. 
In the case $\tau=0.01$, $m(t)$ shows ballistic behavior ($m(t) \sim t^2$) which 
is due to "motional narrowing" caused by the short-time correlation.
On the other hand, in the case $\tau=10^5$ $m(t)$ shows typical localization behavior
within this time scale due to Anderson localization.
In the intermediate case $\tau=1$, $m(t)$ shows normal diffusion with Gaussian shape.

\begin{figure} [h]
\begin{center}
\includegraphics[scale=0.5,clip]{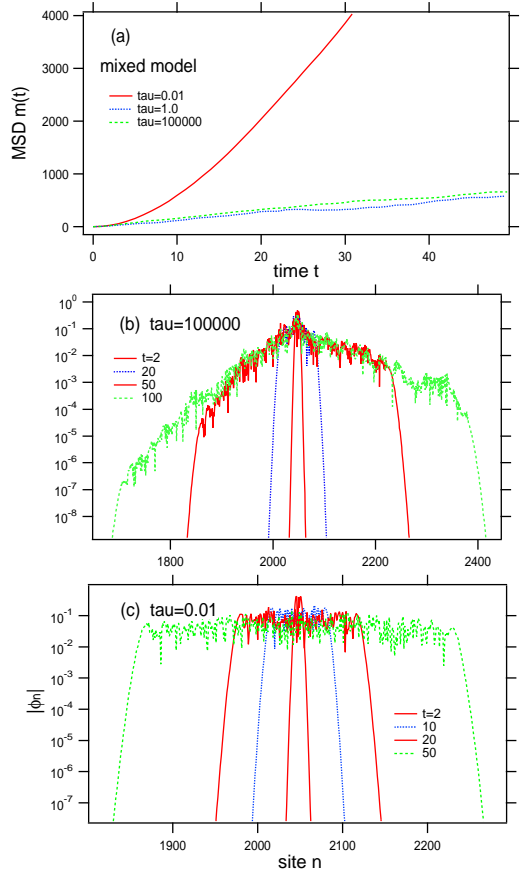}
 \caption{ (a) $m(t)$ at cases with $\tau= 0.001, 1, 100000$ in the mixed model 
with a constant hopping $V_{nn+1}=V_0=1.0$.
Some snapshots of  $| \phi_n(t)| $ at $t=2,20,50,100$ in the case with $\tau=10^5$(b) 
 and $\tau=0.01$(c).
}
\end{center}
\end{figure}

\subsection{Fluctuating hopping term}
Next, we show the numerical results of some cases with fluctuating 
hopping terms $V_{nn+1}(t)$ obeying Eq.(3).
Figure 2 shows some typical time-dependence of the hopping term $V_{12}(t)$ 
for various correlation times $\tau=10^5, 1, 0.01$.
The latter correspond to adiabatic, intermediate and rapid fluctuations, respectively.

Figure 3 displays the time-dependence of the MSD in A-T, G-C and mixed model with $r_{n0}=1.0$
for various correlation times $\tau=10^5, 1, 0.01$. 
Snap-shots of $|\phi_n(t)| $ in A-T model
for various correlation time $\tau=10^5, 1, 0.01$ are shown in Figure 4.
As a result we obtain that the essential behavior does not depend on existence of the fluctuations of the hopping term.
The dynamical behavior is qualitatively similar to the cases discussed in the preceding section with constant $V_{nn+1}=1$.
In the relatively short-correlation case (white-noise limit  $\tau=0.01$), 
the wave packet exhibits ballistic propagation. 
The extent of the spread of the wavepacket in the A-T model is larger than 
that of the G-C model.
On the other hand, in the adiabatic limit ($\tau=10^5$) the wave packet localizes, which 
corresponds to Anderson localization. The localization length is $\ell \sim 20$ sites.
It seems that the localization length of the A-T model is slighty larger than that of the G-C model on this scale.

\begin{figure} [h]
\begin{center}
\includegraphics[scale=0.5,clip]{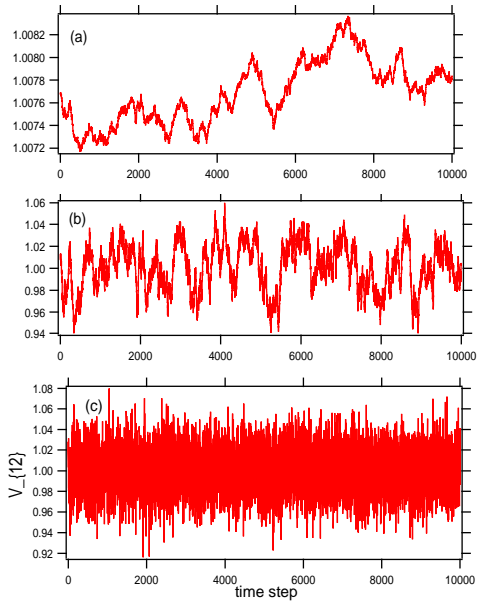}
 \caption{ 
 The time-dependence of the hopping term $V_{12}(t)$ at $n=2$ in A-T model with $r_{n0}=1.0$
for various correlation times  $\tau=10^5$(a),  
$\tau=1.0$(b), $\tau=0.01$(c).
}
\end{center}
\end{figure}

\begin{figure} [h]
\begin{center}
\includegraphics[scale=0.6,clip]{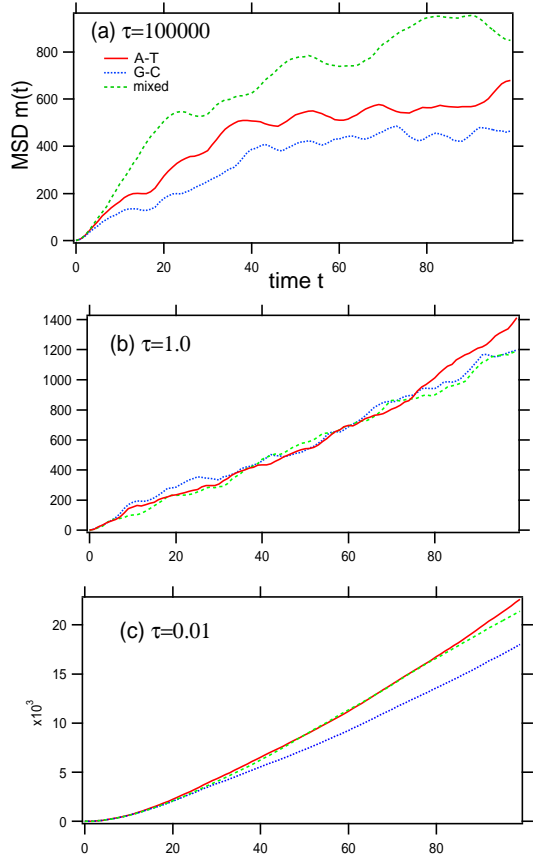}
 \caption{$m(t)$ in A-T, G-C, and mixed models with $r_{n0}=1.0$ and $\tau=10^5$(a),  
$\tau=1.0$(b), $\tau=0.01$(c).
}
\end{center}
\end{figure}

\begin{figure} [h]
\begin{center}
\includegraphics[scale=0.6,clip]{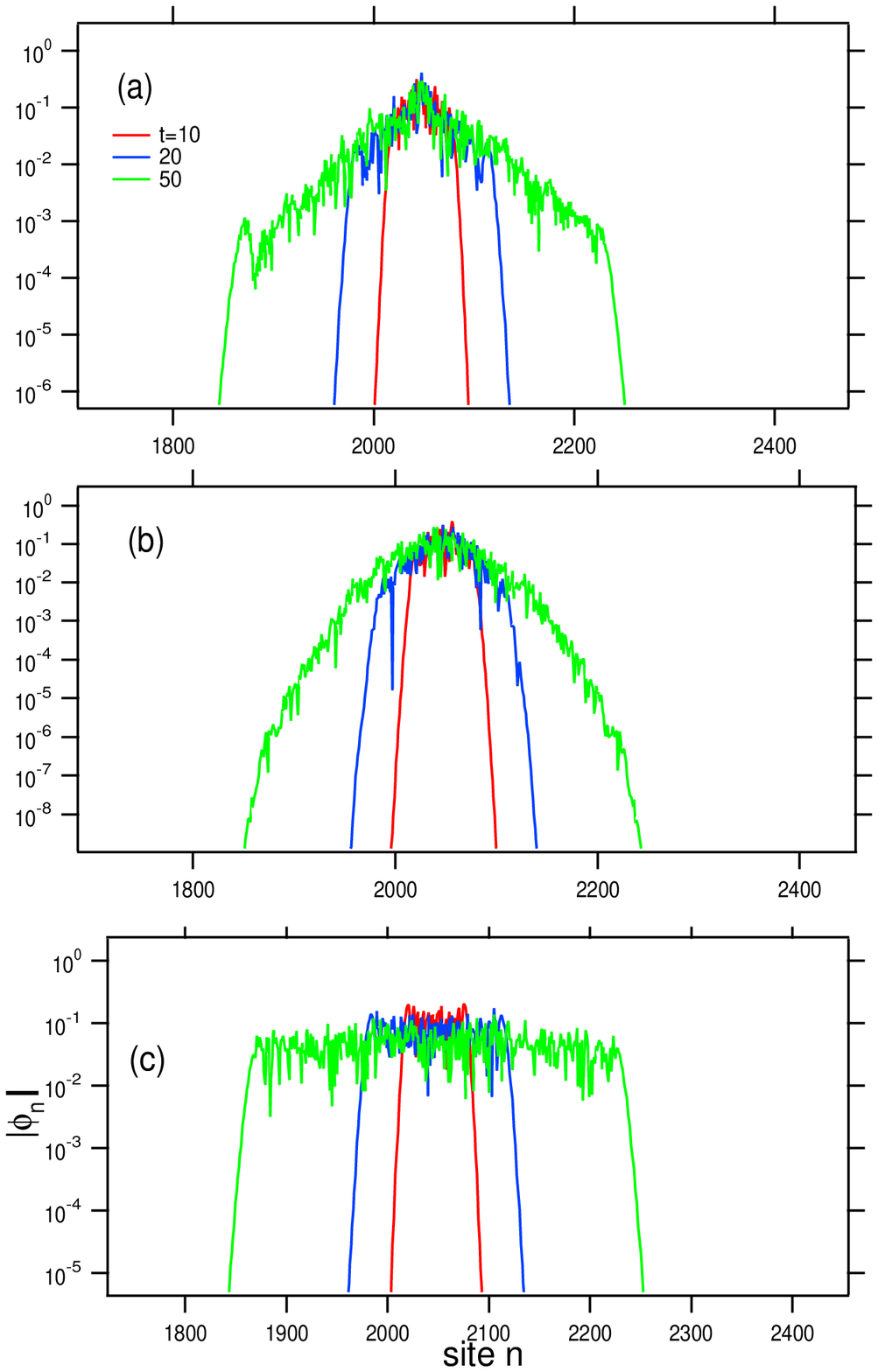}
 \caption{ Some snap shots of $|\phi_n| $ at $t=10,20,50$
in the A-T model with $r_{n0}=1.0$ and  $\tau=0.01$(a), $\tau=1$(b), 
$\tau=10^5$(c).
}
\end{center}
\end{figure}

Next in Fig.5 we show a typical MSD and $P(n,t)$
for the relatively small fluctuation strength $r_{n0}=0.1$  in the mixed model with $\tau=1$. 
As evident from the relation in Eq.(9), 
in comparison to the case with a comparatively large fluctuation amplitude  $r_{n0}=1.0$
(cf. Fig.3(b)) for a smaller amplitude $r_{n0}=0.1$ the quantum diffusion behaves ballistically ($m \sim t^2$)
within the same time scale.
For the $A-T$ and the $G-C$ model we found similar behavior in dependence on the change in the fluctuation strength $r_{n0}$.
\begin{figure} [h]
\begin{center}
\includegraphics[scale=0.5,clip]{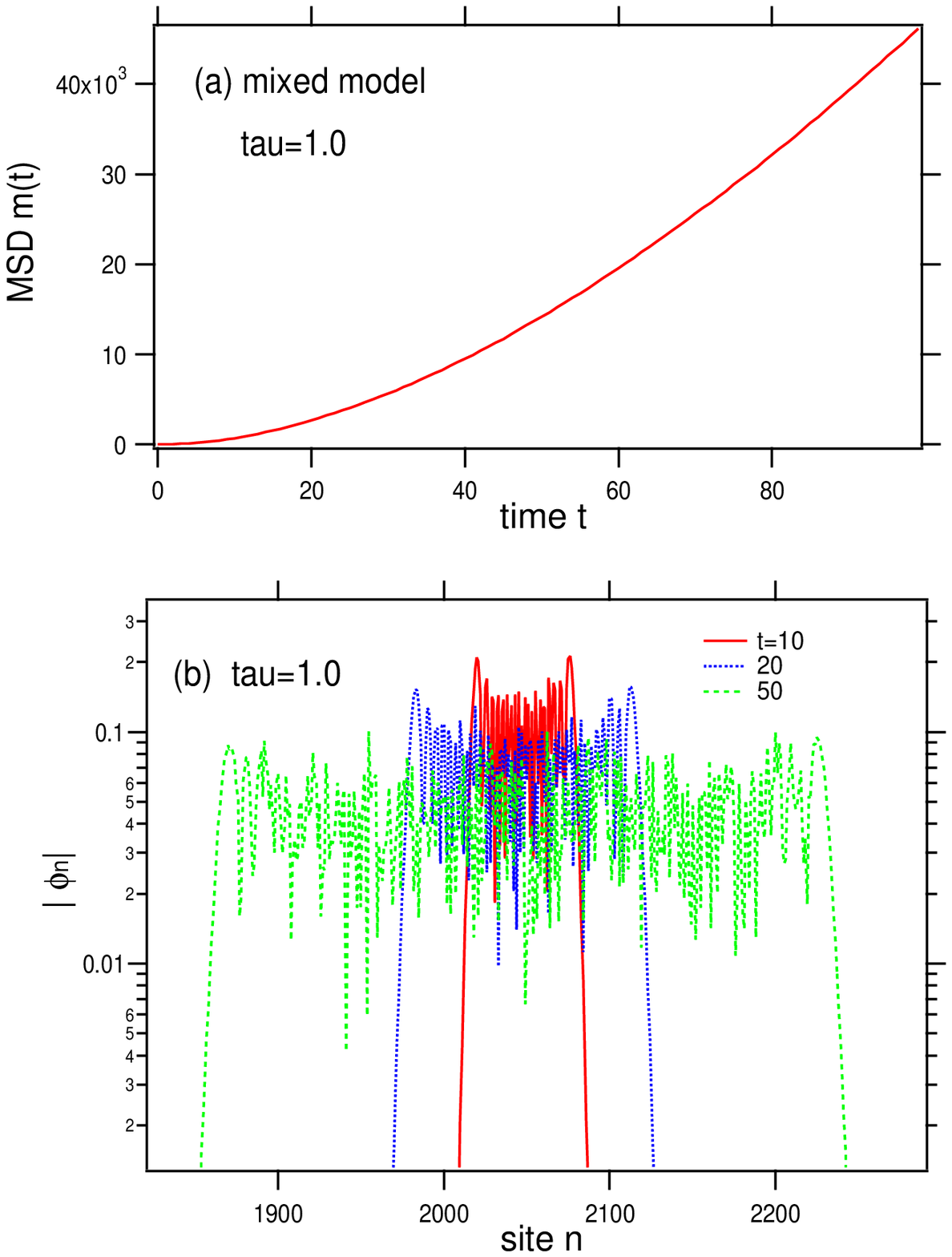}
 \caption{(a) $m(t)$ and (b) $|\phi_n|$  at $t=10,20,50$
 in the mixed model with  $r_{n0}=0.1, \tau=1$.
}
\end{center}
\end{figure}

\subsection{Diffusion rate}
Here we investigate the temporal diffusion rate defined as
\begin{eqnarray}
D(t) = \frac{<m(t)>}{t}
\end{eqnarray}
in the almost diffusive range $0.1 < \tau <10 $, where $<...>$ expresses the average over several samples.
If the motion of the wave packet is diffusive the diffusion rate is supposed to attain a constant value. 
Figure 6 shows the diffusion rate as a function of 
$1/\tau$ for various fluctuation amplitudes $r_{n0}$ 
in A-T and G-C models.  For the estimate of the diffusion rate we used a time interval $t=10^5 \delta t$. where the error of the diffusion 
rate is less than ten percent.  $D(t)$ approaches zero for adiabatic limit ($\tau >> 1$) due to 
Anderson localization, while $D(t)$ goes to infinity 
for motional narrowing case ($\tau << 1$).  
As indicated in Sect.3, increase of $r_{n0}$ means increase of
temperature.  It follows that the diffusion rate in the  A-T model is larger than that in 
G-C model at the relatively low temperature ($r_{n0}=1$).  However, at higher temperatures ($r_{n0}=3,5$)
the two models exhibit virtually equal behavior in this correlation-time range.
Furthermore, in the high-temperature regime there seem to be no pronounced alterations of the 
diffusion coefficient as a function of $\tau^{-1}$.

\begin{figure} [h]
\begin{center}
\includegraphics[scale=0.7,clip]{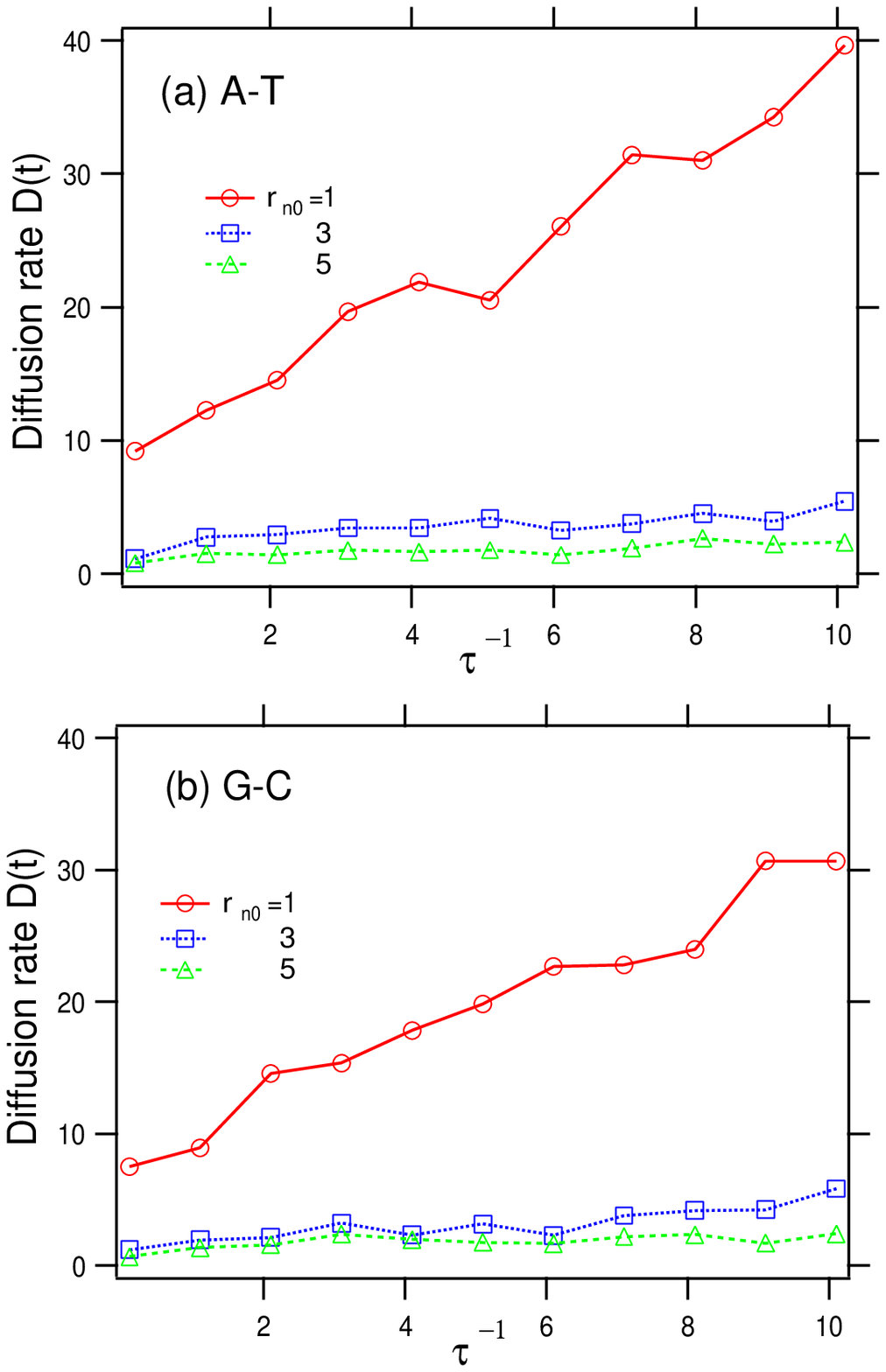}
 \caption{Diffusion rate $D(t)$ as a function of $\tau^{-1}$ 
for several fluctuation strengths $r_{n0}=1, 3, 5$
at A-T model(a) and G-C model(b), respectively
}
\end{center}
\end{figure}

\subsection{Comparison with experimental results}
\label{sec04}
As mentioned in the introduction, in experiments concerning  
the temperature effect on the single-molecule conductance of double-stranded
oligo-DNA with homogeneous base-pairs Zalinge {\it et al} \cite{zalinge06}  found 
temperature-independent ballistic conductance.
Furthermore, it has been shown that $(dG)_{15}-(dC)_{15} $ is a better conductor than 
$(dA)_{15}-(dT)_{15} $ in the conducting experiments, just in accordance 
with the earlier experimental and theoretical findings \cite{hennig04}.
The work \cite{zalinge06} explained the observed temperature independence by 
deactivation of the acoustic modes of DNA at room and higher temperatures.

It is interesting to recall in this context an earlier work \cite{silbey00}, 
where almost  temperature independent photoinjected electron/hole mobility
 has been calculated on the  basis of the stochastic Haken-Strobl-Reinecker (HSR) model 
for discotic liquid crystals. Unlike  in our model, the HSR Hamiltonian 
used in \cite{silbey00} treats all the vibrational and  lattice modes 
as a source of fluctuations of the parameters of the  tight-binding model, 
without singling out any  mode responsible for polaron formation. 
As a result, the HSR mobility temperature independence  shows up only 
at small fluctuations of the molecules around their equilibrium positions. This is not the case in our present model.

We have to pay attention to relatively short-time behavior and/or small spread regime
($\surd m  \sim 15$), as we compare the experimental result with our numerical 
results. 
Figure 7 shows the MSD in the A-T model with $\tau=0.01$ and $\tau=0.0001$ respectively 
in which cases ballistic propagation is observed due to the motional narrowing.
In the short-time regime 
the time dependence of the MSD shows relatively weak $\tau-$dependence 
once the motional narrowing has affected the diffusive behavior. 
Therefore, the temperature-independence may be caused by the motional narrowing 
and the finite-size effect in the experiment.

Figure 8 shows the short-time behavior of the cases $\tau=1$ and $\tau=0.01$ depicted on a larger time scale in Fig.3. 
It follows that in the short-time behavior the  $(dG)_{15}-(dC)_{15}$ case is more diffusive 
than its $(dA)_{15}-(dT)_{15}$ counterpart within the range fro which the spread of the wavepacket is 
$\surd m  \sim 15$.

\begin{figure} [h]
\begin{center}
\includegraphics[scale=0.6,clip]{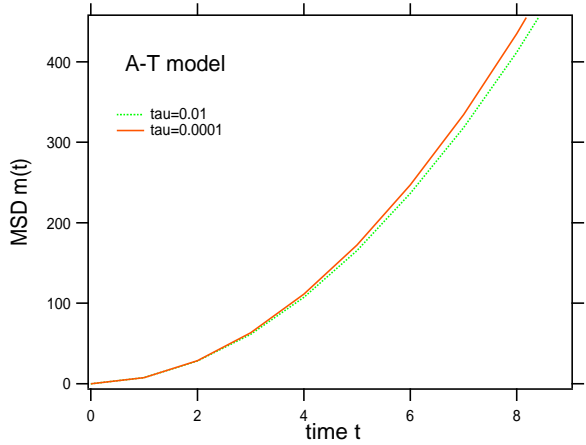}
 \caption{Short-time behavior of $m(t)$ in the A-T model with $\tau=0.01, 0.0001$.  
}
\end{center}
\end{figure}

\begin{figure} [h]
\begin{center}
\includegraphics[scale=0.7,clip]{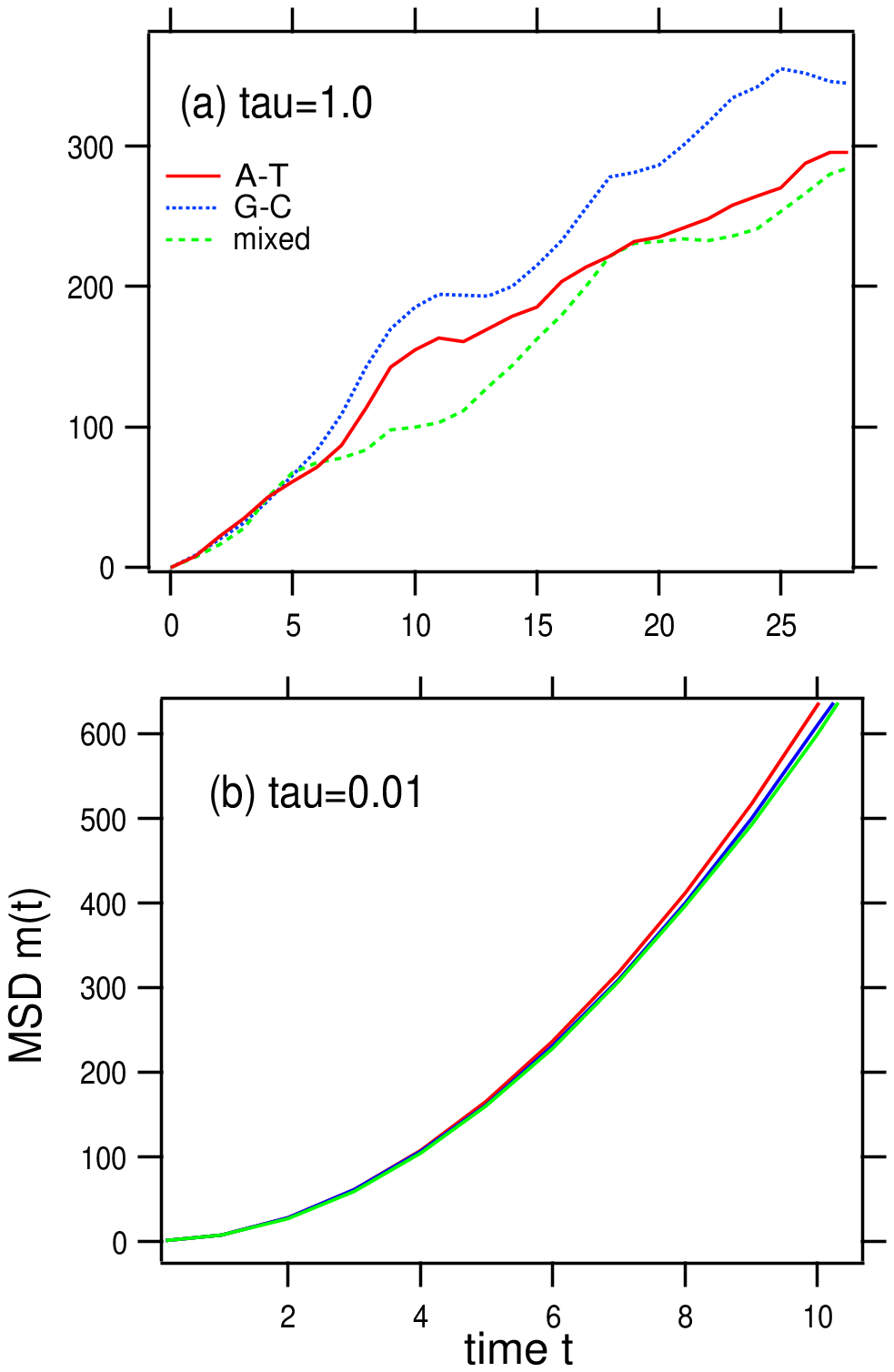}
 \caption{Short-time behavior of $m(t)$ of A-T, G-C, and mixed models 
with $r_{n0}=1.0$ at $\tau=1$(a) and 
 $\tau=0.01$(b).
}
\end{center}
\end{figure}


\section{Summary and discussion}
\label{sec05}
We have numerically investigated quantum diffusion of an electron in the Hennig
 model of poly(dG)-poly(dC) and poly(dA)-poly(dT) DNA
polymers with fluctuations of the parameters caused by colored noise.
In the model the decay time of the correlation can control the spread of the 
electronic wavepacket. 
It has been  shown that in a motional narrowing regime the averaging over fluctuations 
causes ballistic propagation of the wavepacket
and in the adiabatic regime the electronic states are strongly localized.
It has been demonstrated that the motional narrowing affects the
localization in the poly(dG)-poly(dC) and poly(dA)-poly(dT) DNA polymers.
In either model temperature-dependence becomes virtually suppressed when the motion of the 
wave packet is characterized by ballistic propagation. 

We have also investigated the temporal diffusion rate 
in the almost diffusive range.
It has been found that the diffusion rate of the  A-T model is larger than that of 
the G-C model at comparatively low temperatures. Interestingly
for relatively high temperatures in the diffusive range of the wavepacket motion 
the difference between the two DNA systems gets smaller.

Furthermore we commented on the 
relation between our numerical results and the experimental ones.
It was shown that 
 in the short-time behavior the significant difference of the spread of wavepackets 
 does not exist
 between $(dA)_{15}-(dT)_{15} $ and  $(dG)_{15}-(dC)_{15} $ except for the localization cases.

In the present report 
we used periodic sequences, i.e. constant values of $E_0$, $V_0$,  as static parts of the on-site and 
hopping terms for poly(dG)-poly(dC) and
poly(dA)-poly(dT) DNA polymers, respectively. This also includes the mixed model.
Then the motional narrowing for dynamical disorder makes
 the time-evolution of the 
wavepacket ballistic.  
However, it should be remarked that  
motional narrowing strongly localizes the 
wavepacket if we use disordered sequence for the static parts of 
$E_n$ and/or $V_{nn+1}$.

\appendix
\section{Generation of colored noise}
In this appendix, 
we give an algorithm in order to generate the Gaussian colored noise $r_n(m\delta t)$
at the $m$th time step \cite{bartosch01}.
First, let us assume Gaussian random numbers $z_n(m)$ with zero mean and unit variance
at the time $t_m=m \delta t$. Then the stochastic sequence with the colored correlation 
is obtained by the following recursion: 
\begin{eqnarray}
r_{n}(0)  &=&  r_{n0} z_n(0) \\
r_{n}(m)  &=& \rho_m r_n(m-1) + \surd (1-\rho_m^2)  \{ r_{n0} z_n(m) \} , 
\end{eqnarray}
where $\rho_0=0$, $\rho_m = \exp (-|t_m-t_{m-1}|/\tau )$.  
Note that index $n$ denotes the site $n$ and $m$ the time step $m$. 
In our numerical calculation, $\rho_m =const.(= \exp (-\delta t/\tau))$.  
We used the algorithm independently for each site $n$.


It should be noted, that there is another algorithm using the power spectrum 
of the stochastic process,  which is applicable to a lot of types
 of the correlated sequence 
\cite{silbey00,haken73,sumi77,inaba81,ezaki91,noba01,kampen89,moura98,deych03,adame03,cohen91}.  


More recently, it has been shown that 
motional narrowing due to large thermal fluctuation
effects the coherence of relaxation dynamics such as 
spin relaxation in semiconductors \cite{berthelot06}
or vibrational dephasing in a spin-Peierls system 
with lattice fluctuations \cite{onishi03}.

\section*{Acknowledgments}
We would like to thank Professor Juan F.R. Archilla
for discussion in first stage of this work.
H.Y. would like to thank Shuichi Kinosita for 
 sending me some related papers. 


\end{document}